\documentclass[aip,reprint]{revtex4-1}

\usepackage{amsmath}
\usepackage{amssymb}
\usepackage{amsfonts}
\usepackage{bm}
\usepackage{graphicx}
\usepackage{comment}
\usepackage{color}
\usepackage{siunitx}

\usepackage[colorlinks=true,allcolors=blue,bookmarks=true,bookmarksnumbered=true]{hyperref}

\newcommand{\kB}{k_\mathrm{B}}
\newcommand*{\tran}{^{\mkern-1.5mu\mathsf{T}}}

\begin{document}


\title{High-Accuracy Determination of {P}aul-Trap Stability Parameters for Electric-Quadrupole-Shift Prediction} 



\author{T. Lindvall}
\email[]{thomas.lindvall@vtt.fi}
\affiliation{VTT Technical Research Centre of Finland Ltd, National Metrology Institute VTT MIKES, P.O.\ Box 1000, FI-02044 VTT, Finland}

\author{K. J. Hanhij\"arvi}
\affiliation{VTT Technical Research Centre of Finland Ltd, National Metrology Institute VTT MIKES, P.O.\ Box 1000, FI-02044 VTT, Finland}

\author{T. Fordell}
\affiliation{VTT Technical Research Centre of Finland Ltd, National Metrology Institute VTT MIKES, P.O.\ Box 1000, FI-02044 VTT, Finland}

\author{A. E. Wallin}
\affiliation{VTT Technical Research Centre of Finland Ltd, National Metrology Institute VTT MIKES, P.O.\ Box 1000, FI-02044 VTT, Finland}


\date{\today}

\begin{abstract}
The motion of an ion in a radiofrequency (rf) Paul trap is described by the Mathieu equation and the associated stability parameters that are proportional to the rf and dc electric field gradients. Here, a higher-order, iterative method to accurately solve the stability parameters from measured secular frequencies is presented. 
It is then used to characterize an endcap trap by showing that the trap's radial asymmetry is dominated by the dc field gradients and by measuring the relation between the applied voltages and the gradients. The results are shown to be in good agreement with an electrostatic finite-element-method simulation of the trap. Furthermore, a method to determine the direction of the radial trap axes using a `tickler' voltage is presented and the temperature dependence of the rf voltage is discussed. 
As an application for optical ion clocks, the method is used to predict and minimize the electric quadrupole shift (EQS) using the applied dc voltages. Finally, a lower limit of 1070 for the cancellation factor of the Zeeman-averaging EQS cancellation method is determined in an interleaved low/high EQS clock measurement. This reduces the EQS uncertainty of our $^{88}$Sr$^+$ optical clock to ${\lesssim} 1\times 10^{-19}$ in fractional frequency units.
\end{abstract}


\maketitle 


\section{Introduction}

Paul traps or radiofrequency (rf) ion traps \cite{Paul1990a} are widely used in physics, quantum technologies, and precision metrology for applications such as 
quantum information processing and computing,\cite{Haffner2008a,Wineland2011a,Bruzewicz2019a} mass spectrometry,\cite{March2009a,Nolting2017a} and optical atomic clocks.\cite{Margolis2009a,Poli2013a,Ludlow2015b} Many applications require precise control and characterization of the motion of the ion(s), which at Doppler-cooling temperatures, where the motion can be treated classically, is governed by the Mathieu equation.\cite{McLachlan-TAMF} 
Its lowest-order solution  is simple and is, for illustrative purposes, used widely enough that its limitations can easily be forgotten.
In this article, it is demonstrated that the lowest-order expression for the secular frequencies is not generally sufficient for solving the Mathieu stability parameters. Instead, a higher-order, iterative method that gives accurate parameter values after a few iterations is presented. It is then experimentally demonstrated by characterizing an endcap trap and applied to minimizing the electric quadrupole shift (EQS) in a $^{88}$Sr$^+$ single-ion optical clock.

This paper is organized as follows: Section \ref{sec:theory} reviews the Paul trap and then presents a high-accuracy method to solve the stability parameters, including a way to determine whether the radial asymmetry is dominated by rf or dc asymmetry. Section \ref{sec:exp} describes the experimental setup, how to measure secular frequencies using a `tickler' voltage, and demonstrates a way to determine the direction of the radial trap axes. It is then shown that the radial asymmetry of our trap is dominated by the dc field gradients and the relation between the applied voltages and the gradients is obtained experimentally. Also the temperature dependence of the trap rf voltage is discussed. Section \ref{sec:FEM} presents a finite-element-model simulation of the trap that gives the electric field strength and gradient as functions of the applied voltages, in good agreement with the experimental results. The observed constant gradient is discussed in Sec.~\ref{sec:gradEconst}. In Sec~\ref{sec:EQS}, a simple relation between the 
EQS and the $a_i$ trap parameters is derived and it is shown how the EQS can be predicted and minimized. Finally, a lower limit for the cancellation factor of the Zeeman-averaging EQS cancellation method \cite{Dube2005a} is measured.

\section{Theory \label{sec:theory}}

\subsection{The Paul trap}

The electric potential of an ideal spherical Paul trap, where the electrodes are infinite hyperboloids, is \cite{Wineland1983a}
\begin{equation}
\phi(r,z) = \frac{U_0 + V_0 \cos{\Omega t}}{r_0^2 + 2 z_0^2} (r^2 - 2 z^2).
\end{equation}
Here $U_0$ and $V_0$ are the applied dc and rf voltages, $\Omega$ is the rf frequency, $r_0$ is the radius of the ring electrode, and $2 z_0$ is the separation of the endcap electrodes. Practical trap geometries deviate significantly from the ideal one, and for the endcap trap design,\cite{Schrama1993a} which is frequently used in single-ion optical clocks for its superior optical access, the radius $r_0$ is not well defined. In this case, one can use the relation $r_0 = \sqrt{2} z_0$ of the ideal geometry and introduce the trap efficiency $\eta$ (or its inverse, the voltage loss\cite{Schrama1993a}) to describe the fraction of the applied voltage that contributes to the quadrupole component. In addition, the radial asymmetry is often accounted for by an asymmetry parameter $\epsilon$,
\begin{eqnarray} \label{eq:phi2}
\phi(x,y,z) &=& \frac{\eta_U U_0 + \eta_V V_0 \cos{\Omega t}}{4 z_0^2} \nonumber \\
&& \times \left[ (1-\epsilon) x^2 + (1+\epsilon) y^2 - 2 z^2\right].
\end{eqnarray}
Note that perfect radial symmetry is not desirable, as it makes Doppler cooling using a single laser beam impossible. \cite{Javanainen1980e}
Often the rf is applied to the endcap (inner) trap electrodes while the dc is applied to the shield (outer) electrodes, which results in slightly different trap efficiencies $\eta_V$ and $\eta_U$, respectively.

Equation~\eqref{eq:phi2} assumes that the radial asymmetry is due to geometric imperfections of the trap electrodes alone. However, the bias (trim) electrodes used to minimize excess micromotion (EMM) by cancelling stray electric fields can also contribute significant dc field gradients, as can patch and contact potentials. 
If the radial asymmetry of both the rf field and the dc field are significant and the two fields have different symmetry axes, there is no coordinate system where the radial equation of motion decouples into two one-dimensional Mathieu equations.\cite{Shaikh2012a} 
This case is discussed in Appendix~\ref{sec:diff-axes}.
In the following, it is assumed that a set of well defined principal axes exists and a method that shows whether the radial asymmetry is dominated by the rf or dc field is presented.

With decoupled axes, the potential can be written as
\begin{equation} \label{eq:phiaq}
\phi(\mathbf{x}) = \frac{m \Omega^2}{8 q} \sum_i \left(a_i - 2q_i \cos{\Omega t} \right) x_i^2,
\end{equation}
where $m$ is the ion mass and $q$ the ion charge. 
The stability parameters $a_i$ and $q_i$ are proportional to the gradient of the dc and rf electric field, respectively,
\begin{equation} \label{eq:gradE}
\frac{d}{d x_i}(E_i + E_{\mathrm{rf},i}) = -\frac{d^2 \phi}{dx_i^2} = -\frac{m \Omega^2}{4 q} \left(a_i - 2q_i \cos{\Omega t} \right).
\end{equation}
Equations \eqref{eq:phiaq} and \eqref{eq:phi2} yield $q_z = 2\eta_V q V_0/(m \Omega^2 z_0^2)$ and
$q_{x,y} = -(1\mp \epsilon) q_z/2$. Similar expressions exist for the $a_i$ parameters, but to allow for dc gradients not created by the applied voltages, these are are kept 
as `free' parameters, constrained only by $\sum_i a_i = 0$
as required by the Laplace equation.

The equations of motion of the ion can then be written in the form of the Mathieu equation,
\begin{equation} \label{eq:mathieu}
\frac{d^2 x_i}{d\tau^2} + \left(a_i - 2q_i \cos{2\tau}\right) x_i = 0,
\end{equation}
where $\tau = \Omega t/2$. The stable solution to Eq.~\eqref{eq:mathieu} can be written as \cite{McLachlan-TAMF}
\begin{eqnarray} \label{eq:x-exact}
x_i(t) &=& C \sum_{n=-\infty}^{\infty} c_{2n} \cos{\left[ (2n+\beta_i)\tau+\phi_i \right]} \nonumber \\
&=&  C \sum_{n=-\infty}^{\infty} c_{2n} \cos{\left[ (n\Omega+\omega_i)t+\phi_i \right]}.
\end{eqnarray}
The parameter $\beta_i$, which defines the secular frequency $\omega_i = \beta_i\Omega/2$, can be solved numerically and the coefficients $c_{2n}$ can then be solved to the desired order using recursion relations.\cite{McLachlan-TAMF} The phase of the secular motion, $\phi_i$, depends on the initial conditions and the normalization constant $C$ is obtained from the temperature (kinetic energy) of the ion using $\kB T = m\langle (dx_i(t)/dt)^2\rangle /2$.

To lowest order in $q_i$ and $a_i$, Eq.~\eqref{eq:x-exact} becomes
\begin{equation}
x_i(t) \approx C_1 \cos{(\omega_i t + \phi_i)} \left( 1- \frac{q_i}{2} \cos{\Omega t} \right),
\end{equation}
with the secular frequency given by
\begin{equation} \label{eq:secular1}
\omega_i = \frac{\Omega}{2}\beta_i \approx \frac{\Omega}{2} \sqrt{a_i + q_i^2/2}.
\end{equation}
This solution is commonly used as it is simple and demonstrates the existence of secular motion and intrinsic micromotion. It is, however, not always sufficient as will be shown below.

\subsection{Determining the stability parameters \label{sec:determine}}

If $q_i$ and $a_i$ are known and $a_i\ll q_i^2\ll 1$, Eq.~\eqref{eq:secular1} gives a reasonable estimate for the secular frequency. However, one often wants to solve the $q_i$ and $a_i$ parameters from measured secular frequencies instead. Using Eq.~\eqref{eq:secular1} for this, the relative error of the $a_i$ values can be huge, as typical $a_i$ values are much smaller than the neglected $q_i^4$ and even $q_i^6$ terms. To obtain more accurate $a_i$ and $q_i$ parameters, we therefore use the following approximation,
\cite{McLachlan-TAMF}
\begin{eqnarray}
\beta_i^2 &\approx& a_i - \frac{a_i-1}{2(a_i-1)^2-q_i^2} q_i^2 - \frac{5a_i+7}{32(a_i-1)^3(a_i-4)} q_i^4 \nonumber \\
&& - \frac{9a_i^2 + 58 a_i +29}{64(a_i-1)^5(a_i-4)(a_i-9)} q_i^6,
\end{eqnarray}
which we expand to order $a_i^1$ and $q_i^6$,
\begin{eqnarray} \label{eq:betaseries}
\beta_i^2 &\approx& a_i + \left(\frac{1}{2}+\frac{a_i}{2}\right)q_i^2
+ \left(\frac{25}{128}+\frac{273}{512}a_i\right)q_i^4 \nonumber \\
&&+ \left(\frac{317}{2304}+\frac{59525}{82944}a_i\right)q_i^6.
\end{eqnarray}
Note that $a_i^2$ does not occur by itself: the lowest-order neglected term is $\mathcal{O}(a_i^2 q_i^2)$.

By combining expression \eqref{eq:betaseries} for $\beta_x^2$, $\beta_y^2$, and $\beta_z^2$ in suitable ways, all the trap parameters can iteratively be solved to good accuracy.
First, by forming the sum $\sum_i \beta_i^2$, the $a_i$ terms in \eqref{eq:betaseries} cancel as $\sum_i a_i = 0$ and we obtain a cubic equation for $q_z$,
\begin{equation} \label{eq:qzcubic}
c_6 q_z^6 + c_4 q_z^4 + c_2 q_z^2 - \sum_i \beta_i^2 = 0,
\end{equation}
where
\begin{widetext}
\begin{subequations} \label{eq:cubic-cs}
\begin{eqnarray}
c_6 &=& \frac{317}{2304} \left[ 1 + \left(\frac{1-\epsilon}{2}\right)^6 + \left(\frac{1+\epsilon}{2}\right)^6 \right]
+  \frac{59525}{82944} \left[ a_z + a_x \left(\frac{1-\epsilon}{2}\right)^6 + a_y\left(\frac{1+\epsilon}{2}\right)^6 \right] \\
c_4 &=& \frac{25}{128} \left[ 1 + \left(\frac{1-\epsilon}{2}\right)^4 + \left(\frac{1+\epsilon}{2}\right)^4 \right]
+  \frac{273}{512} \left[ a_z + a_x \left(\frac{1-\epsilon}{2}\right)^4 + a_y\left(\frac{1+\epsilon}{2}\right)^4 \right] \\
c_2 &=& \frac{1}{2} \left[ 1 + \left(\frac{1-\epsilon}{2}\right)^2 + \left(\frac{1+\epsilon}{2}\right)^2 \right]
+  \frac{1}{2} \left[ a_z + a_x \left(\frac{1-\epsilon}{2}\right)^2 + a_y\left(\frac{1+\epsilon}{2}\right)^2 \right].
\end{eqnarray}
\end{subequations}
\end{widetext}
This equation has only one real root and can be solved using standard formulae for cubic equations. To obtain a first iterative value for $q_z$, one must neglect the $a_i$ parameters as well as $\mathcal{O}(\epsilon^2)$. A first value for $a_z$ can then be solved from \eqref{eq:betaseries},
\begin{equation} \label{eq:az}
a_z = \frac{ \beta_z^2 - \frac{q_z^2}{2} - \frac{25}{128}q_z^4 - \frac{317}{2304}q_z^6}
{1 + \frac{q_z^2}{2} + \frac{273}{512}q_z^4 + \frac{59525}{82944}q_z^6}.
\end{equation}

If the rf field is not radially symmetric, $\epsilon \neq 0$, there are four unknown parameters, $q_z$, $\epsilon$, and two independent $a_i$ parameters, so three measured secular frequencies are not sufficient to solve them all. Taking $\epsilon$ to first order, the radial asymmetry is described by the relation
\begin{equation} \label{eq:betaradial}
\beta_y^2 - \beta_x^2 = (a_y - a_x) f_1(q_z) + \epsilon f_2(q_z, a_z),
\end{equation}
where
\begin{subequations}
\begin{eqnarray}
f_1(q_z) &=& 1 + \frac{q_z^2}{8} + \frac{273}{2^{13}}q_z^4 + \frac{59525}{81\times 2^{16}}q_z^6, \\
f_2(q_z,a_z) &=& \left(\frac{1}{2}-\frac{a_z}{4}\right)q_z^2 + \left(\frac{25}{256}-\frac{273}{2^{11}}a_z\right)q_z^4 \nonumber \\
&&+ \left(\frac{317}{3\times 2^{12}}-\frac{178575}{81\times 2^{15}}a_z\right)q_i^6.
\end{eqnarray}
\end{subequations}
Here $a_x + a_y = -a_z$ was used.
By measuring the secular frequencies for different rf voltages $V_0$, i.e., different $q_z$, both $(a_y - a_x)$ and $\epsilon$ can be solved by multiple linear regression of  Eq.~\eqref{eq:betaradial} with $f_1$ and $f_2$ as the independent variables.
The obtained $a_x$, $a_y$, and $\epsilon$ values can then be substituted into Eqs.~\eqref{eq:cubic-cs} and new trap parameter values can be solved from Eqs.~(\ref{eq:qzcubic}, \ref{eq:az}--\ref{eq:betaradial}). For typical, small $a_i$ and $\epsilon$ values, a few iterations are enough to obtain an accuracy limited by the approximation \eqref{eq:betaseries}, which neglects terms of order $q_z^8$ and higher.

If the analysis above shows that the asymmetry of the rf field is negligible, the analysis is significantly simplified and a single set of secular frequencies is sufficient to solve the trap parameters. In this case, $q_z$ is solved from \eqref{eq:qzcubic} and the radial symmetry gives $q_x = q_y = -q_z/2$. All $a_i$ parameters can then be solved using expressions like Eq.~\eqref{eq:az}.

The accuracy of the iterative approach can be determined by choosing stability parameters $q_i$ and $a_i$, calculating the corresponding $\beta_i$ values to high order, and then calculating approximative $q_i$ and $a_i$ values from these. This shows that the error of the estimated values is of order $\mathcal{O}(q_i^8,\, a_i^2 q_i^2)$, as expected from the series expansion in Eq.~ \eqref{eq:betaseries}. For the typical values $q_z = 0.4$ and $|a_i| \leq 0.01$, the error of $q_z$ is ${\sim} 10^{-4}$ and that of the $a_i$ parameters ${\sim} 10^{-5}$.

\section{Experimental \label{sec:exp}}

\subsection{Endcap trap apparatus}

\begin{figure}[tb]
\centering
\includegraphics[width=1\columnwidth]{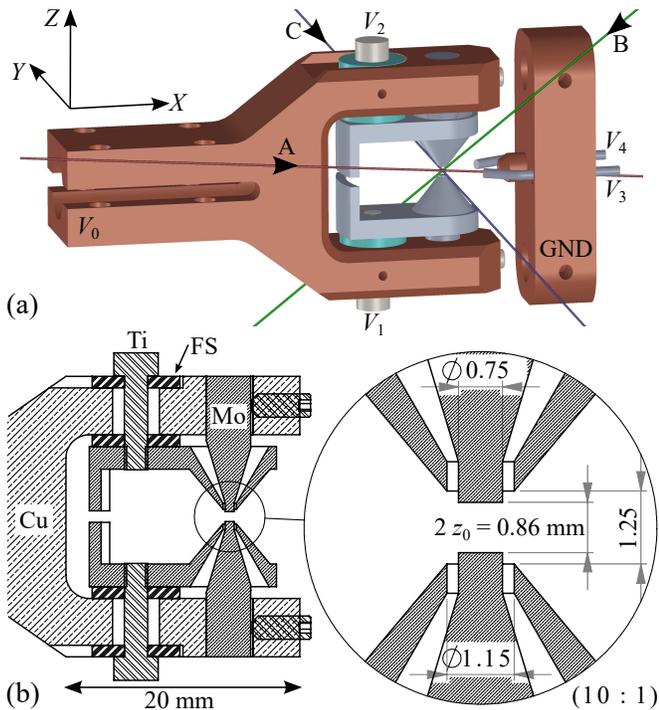}\\%
\caption{(a) Trap 3D drawing. Laser beams A, B, and C cross at the position of the ion; arrowheads indicate direction of propagation. The chamber coordinate system $XYZ$ is shown offset from the origin at the ion for clarity. Labels indicate how the rf voltage $V_0$ and the dc bias voltages $V_j$ ($j=1\ldots 4$) are connected. (b) Section view. The different hatches indicate the materials: Cu--copper, Mo--molybdenum, Ti--titanium, FS--fused silica. The inset shows the electrode geometry enlarged.
\label{fig:trap}}
\end{figure}

The endcap trap used in this work is similar to the one in \citet{Nisbet-Jones2016a}, but with slightly different dimensions, see Fig.~\ref{fig:trap}. The rf voltage $V_0$ is symmetrically supplied to the molybdenum endcap electrodes through the copper trap body, which is mounted directly to the 6.4-mm copper rod of the rf vacuum feedthrough. The shield electrodes, also from molybdenum, are insulated from the trap body by fused silica spacers and are connected to separate electrical feedtroughs in the same flange as the rf feedthrough, with low-pass filters immediately on the atmosphere side. The trap dc voltage is applied in common mode to the two shield electrodes, $U_0 = V_{12,\mathrm{cm}} = (V_2 + V_1)/2$, whereas a differential voltage $V_{12,\mathrm{diff}} = (V_2 - V_1)/2$ is used to compensate stray fields along the trap axis $Z$. In the radial plane, two bias electrodes at $\pm30^\circ$ angles to the $X$ axis
are used for stray field compensation. The common-mode, $V_{34,\mathrm{cm}} = (V_4 + V_3)/2$, and differential, $V_{34,\mathrm{diff}} = (V_4 - V_3)/2$, voltages are used to control the electric field along the $X$ and $Y$ axes, respectively.

The rf is supplied to the trap by a tunable helical resonator, see 
Appendix~\ref{sec:helres}, operating at the $^{88}$Sr$^+$ magic frequency of 14.4\;MHz, \cite{Dube2014a} where the micromotion shifts cancel.
An rf power of 250\;mW into the resonator results in a trap voltage (amplitude) of $V_0 = 350$\;V. 
The compact vacuum chamber (internal volume ${\sim}0.3\;\mathrm{dm}^3$) is made of aluminum and is pumped by a SAES Getters NEXTorr D-200 pump, which
results in a pressure in the low $10^{-11}$\;mbar range at the position of the ion.

The trap is loaded from a beam of neutral Sr atoms from a commercially available oven (Alvatec Alvasource AS-2-Sr-45-F).
Inside the trap volume, the atoms are photoionized in a resonant two-step process  \cite{Brownnutt2007a} using a frequency-doubled single-mode distributed Bragg reflector (DBR) laser at 461\;nm 
and a free-running multimode diode laser at 405\;nm.

The ion is Doppler cooled on the $5s \,^2S_{1/2} \rightarrow 5p\, ^2P_{1/2}$ transition using a frequency-doubled distributed feedback (DFB) laser at 422\;nm. \cite{Fordell2014a} It is frequency stabilized to a rubidium line, detuned from the cooling transition by only 436\;MHz, \cite{Shiner2007a} using saturated absorption spectroscopy. \cite{Sinclair2001a} 
For repumping on the 1092-nm $4d\,^2D_{3/2} \rightarrow 5p\, ^2P_{1/2}$ transition and clock-state clearout on the 1033-nm $4d\,^2D_{5/2} \rightarrow 5p\, ^2P_{3/2}$ transition, unpolarized amplified-spontaneous-emission (ASE) light sources are used. \cite{Lindvall2013a,Fordell2015a}
These require no frequency stabilization and the repumper destabilizes dark states without external polarization scrambling. \cite{Lindvall2012a}

Ion fluorescence is detected using an objective close to the anti-reflection--coated fused silica window of the vacuum system. A $70/30$ beam splitter divides the collected photons between a photon multiplier tube (PMT) and an sCMOS camera.

The interrogation laser for the $5s \,^2S_{1/2} \rightarrow 4d\,^2D_{5/2}$ electric quadrupole (E2) 
clock transition is a commercial 1348-nm external-cavity diode laser (Toptica DL PRO), which is amplified and frequency-doubled to 674\;nm and stabilized to a 30-cm-long ultra-low-expansion (ULE) glass cavity of a design similar to the one in \citet{Hafner2015a} The clock interrogation sequence is similar to \citet{Dube2015a}, except that ground-state state preparation is not implemented. Six independent clock servos are locked to transitions addressing all $4d\,^2D_{5/2}$ sublevels. Their frequencies are then combined pairwise to cancel the linear Zeeman effect and all six to cancel the EQS.\cite{Dube2005a} 

The B beam shown in Fig.~\ref{fig:trap} is the main cooling and interrogation beam. Beams A and C are mainly used for EMM minimization \cite{Berkeland1998a,Keller2015a} and to determine the radial trap axes, see Sec.~\ref{sec:axes}.

The static magnetic field at the position of the ion is controlled using three pairs of coils in approximate Helmholtz configurations. Each pair is centered around one of the chamber axes ($X,Y,Z$) and creates a field in the direction of this axis. The relation between the applied currents and the resulting magnetic field as well as the background magnetic field can be solved by measuring the field magnitude using the Zeeman splitting of the clock transition at different combinations of currents.

\subsection{Measuring secular frequencies \label{sec:meas-sec}}

\begin{figure}[tb]
\centering
\includegraphics[width=0.95\columnwidth]{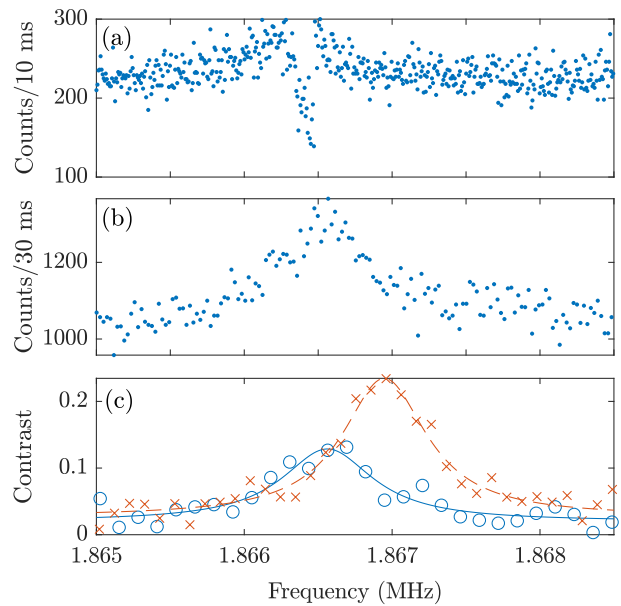}\\%
\caption{Coarse tickler resonances for (a) higher and (b) lower excitation. Vertical axes are PMT counts. (c) Photon correlation tickler resonances with Lorentzian fits for 22\;dB (circles and solid line) and 17\;dB (crosses and dashed line) tickler attenuation. The secular frequency is seen to drift upwards during the course of the measurements.
\label{fig:tickler}}
\end{figure}

The secular frequencies were measured by coupling a `tickler' rf voltage to the upper shield electrode and detecting the change in ion fluorescence when the tickler frequency is scanned over a secular frequency. To quickly find the frequencies, we excite strongly enough that the heating is visible as a clear drop in the fluorescence, as the ion in Doppler-shifted out of resonance, see Fig.~\ref{fig:tickler}(a). This results in asymmetric resonances if the excitation is too strong. By decreasing the excitation level, the resonance is instead seen as an increase in fluorescence at the half-linewidth red detuning of the cooling-laser, as the line is slightly broadened, Fig.~\ref{fig:tickler}(b). An accuracy of ${\lesssim}0.5$\;kHz can be achieved for properly chosen excitation. For higher accuracy, the photon correlation technique, used for micromotion detection, \cite{Berkeland1998a,Keller2015a} is utilized by correlating the fluorescence with the tickler voltage instead of the rf voltage. \cite{Allcock2011PhD} This requires 20\;dB less excitation than the coarse tickler measurement and results in symmetric, Lorentzian resonance curves, as shown in Fig.~\ref{fig:tickler}(c) for two excitation levels. When a Lorentzian fit is used to determine the resonance frequency, the accuracy of this method is mainly limited by the temperature dependence of the secular frequencies, see Sec.~\ref{sec:T-dep}, which is responsible for the drift in Fig.~\ref{fig:tickler}.

Compared to measuring secular frequencies using resolved sideband spectroscopy on the $S_{1/2} \rightarrow D_{5/2}$ clock transition, the main advantage of the tickler methods is the measurement time. In both the coarse and the photon-correlation tickler methods, the tickler voltage from a direct digital synthesizer (DDS) is applied continuously and the frequency is stepped in a staircase-like fashion by phase-continuous tuning of the DDS frequency tuning word. In the coarse tickler scheme, photons are collected for typically 5--30\;ms per frequency, meaning that recording the spectra in Fig.~\ref{fig:tickler}(a) and (b) took 20\;s and 18\;s, respectively. For photon correlation, typically $10^4$ photons are collected at each frequency, which takes 1--2\;s depending on the scattering rate and leads to a total measurement time of the order of 1\;min for the spectra in Fig.~\ref{fig:tickler}(c). Measuring the secular frequencies on the clock transition is significantly slower. In fact, we typically start with a tickler measurement whenever we need to probe the secular sidebands, e.g., for measuring the ion temperature, in order to find them more rapidly.

\subsection{Determining the radial trap axes \label{sec:axes}}

As the tickler voltage is applied to a shield electrode, the resulting field is strong in the $Z$ direction, but significantly weaker in the radial plane. Due to this, exciting the radial modes requires about 12\;dB more excitation than the axial mode. In the radial plane, the L-shaped support parts of the shield electrodes and the trim electrodes, both aligned along the chamber $X$ axis, break the cylindrical symmetry. As the lower-frequency radial mode is more easily excited, this indicates that this trap axis lies closer to the chamber $X$ axis. This is also verified by the camera, where coarse tickler excitation of the lower radial and axial modes is seen as smearing of the ion fluorescence image in the chamber $X$ and $Z$ directions (in the image plane), while exciting the higher radial mode causes hardly any visible effect and must correspond to motion close to the imaging axis (chamber $Y$).

\begin{figure}[tb]
\centering
\includegraphics[width=1\columnwidth]{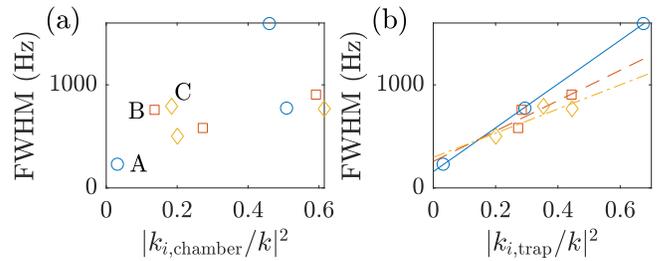}\\%
\caption{(a) Width of photon-correlation secular resonances for three beams, A, B, and C, as a function of the square of the beam projection on the chamber axes. (b) By rotating the coordinate system by $-13^\circ$, the trap axes are found and the widths show a linear dependence for all three beams.
\label{fig:secular-width}}
\end{figure}

Photon-correlation tickler measurements can also be used to determine the direction of the radial trap axes more accurately. The linewidth of a secular resonance depends on the laser cooling rate, \cite{Raab2000a} which is proportional to the square of the projection of the laser beam on the corresponding trap axis. After measuring the width of all three secular resonances using three non-coplanar laser beams, a rotation around the $Z$ axis is applied to make the widths depend linearly on the projection squared for all three beams, see Fig.~\ref{fig:secular-width}.
This indicates that the radial trap axes are rotated by $-13^\circ$ around the $Z$ axis relative to the chamber coordinate system.
In practice, this method requires one laser beam with significantly different projections on the trap axes (the A beam in Figs.~\ref{fig:trap} and \ref{fig:secular-width}). The constant term and the slope of the FWHM lines vary between measurements, possibly depending on the laser-beam intensities, but the linearity criterion works nonetheless and has given consistent results in numerous measurements carried out over 17~months.

\subsection{Experimental stability parameters \label{sec:exp-param}}

Secular frequencies have been measured over a period of 29 months using two different helical resonators operating at $\Omega/2\pi = 16.7$\;MHz and 14.4\;MHz.

\begin{figure}[tb]
\centering
\includegraphics[width=1\columnwidth]{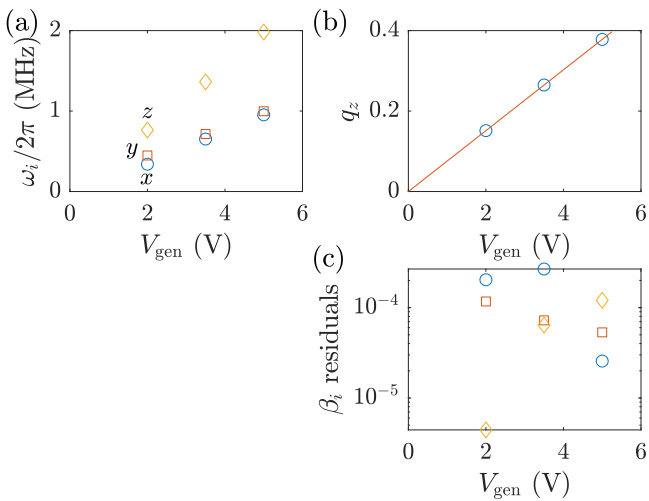}\\%
\caption{(a) Measured secular frequencies of the three motional modes, (b) calculated $q_z$ values and linear fit through the origin, and (c) $\beta_i$ residuals as function of the signal-generator voltage.
\label{fig:vary-V}}
\end{figure}

Figure~\ref{fig:vary-V}(a) shows a measurement of the secular frequencies as a function of the signal-generator voltage amplitude. 
The rf asymmetry parameter obtained by fitting Eq.~\eqref{eq:betaradial} is  $\epsilon = 9\times 10^{-5}$, corresponding to a radial splitting of only $(\omega_y - \omega_x)/2\pi \approx 0.2$\;kHz at radial frequencies around 1\;MHz. On the other hand, the obtained $a_i$ parameters, $(a_x, a_y, a_z) = (-6.5, 9.9, -3.4) \times 10^{-4}$, show a large asymmetry between 
$a_x$ and $a_y$, which explains the measured radial splitting of 47\;kHz.
This is also supported by the trap axis measurements described in Sec.~\ref{sec:axes}: the measured axes do not change even when the rf voltage is lowered by 60\%.  Thus we can assume $\epsilon = 0$ in the following.

The derived $q_z$ parameters are proportional to the signal-generator voltage as expected, see Fig.~\ref{fig:vary-V}(b). The error of the calculated trap parameters is estimated using the fractional $\beta_i$ residuals shown in Fig.~\ref{fig:vary-V}(c), i.e., the relative difference between $\beta_i$ numerically evaluated to high order using the solved parameters and the measured $\beta_i$.
These are identical to the fractional residuals of the secular frequencies $\omega_i$.

\begin{figure}[tb]
\centering
\includegraphics[width=1\columnwidth]{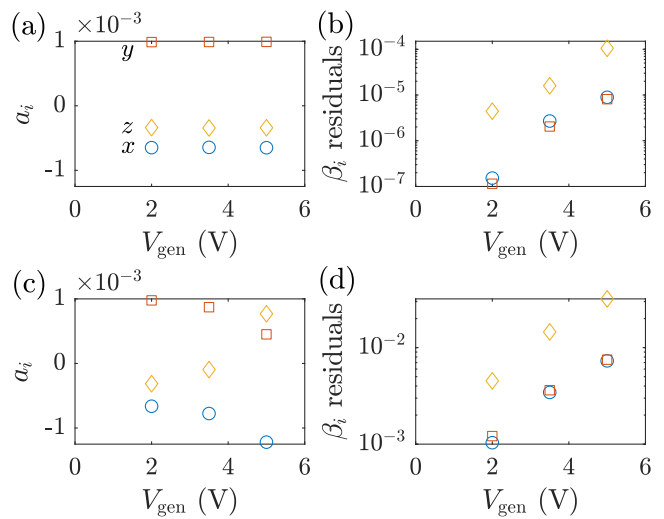}\\%
\caption{$a_i$-parameters (left panels) and $\beta_i$ residuals (right panels) as function of the signal-generator voltage using the high-order, (a) and (b), and lowest-order, (c) and (d), solution.
\label{fig:vary-V-eps0}}
\end{figure}

Assuming $\epsilon = 0$, we can solve the trap parameters from each set of secular frequencies separately. This gives the same $q_z$ values as in Fig.~\ref{fig:vary-V}(b) and the $a_i$ values and $\beta_i$ residuals shown in Fig.~\ref{fig:vary-V-eps0}(a--b). The residuals are lower than in Fig.~\ref{fig:vary-V}, but there are small variations in the $a_i$ values. If we use the mean of these as common values for all rf voltages, the resulting residuals will be of the same order as in Fig.~\ref{fig:vary-V}. These residuals, slightly higher than optimal, are likely due to a drift of the ambient temperature during the measurement, see Sec.~\ref{sec:T-dep}.

For comparison, Fig.~\ref{fig:vary-V-eps0}(c--d) shows the $a_i$ values and $\beta_i$ residuals obtained from the lowest-order expression \eqref{eq:secular1}. For the lowest rf voltage, the $a_i$ values are reasonable, but for the highest voltage, the error in $a_z$ is huge, demonstrating the importance of using the higher-order method. The $\beta_i$ residuals are three orders of magnitude larger for the lowest-order solution.

\begin{figure}[b!]
\centering
\includegraphics[width=1\columnwidth]{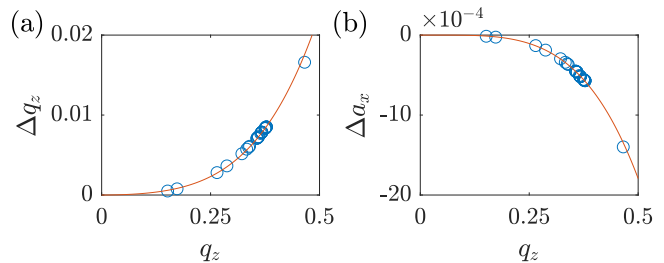}\\%
\caption{Error of lowest-order (a) $q_z$ and (b) $a_x$ values as function of $q_z$ for data with $-1.1\times 10^{-3} < a_x < -0.6\times 10^{-3}$. The fitted curves are of the form $c_2 q_z^2 + c_4 q_z^4$.
\label{fig:all_params}}
\end{figure}

Finally, Fig.~\ref{fig:all_params} plots the error of the lowest-order $q_z$ and $a_x$ parameters relative to the iterative higher-order solution ($a_y$ and $a_z$ show similar differences) for a selection of measurements with $-1.1\times 10^{-3} < a_x < -0.6\times 10^{-3}$. 
For the lowest $q_z$ values of $0.15\ldots 0.17$, the lowest-order solution gives reasonable values, but for larger $q_z$, it fails. For smaller $|a_x|$, it can give the wrong sign. 

\subsection{Electric field gradients as function of dc voltages \label{sec:grad}}

In order to determine the field gradients caused by the applied dc voltages, each voltage was in turn scanned around the value that minimizes the stray fields and the secular frequencies were measured. The scan range for each voltage was chosen to displace the ion from the rf null by ${\sim}\pm1.5\;\mu$m, which is small compared to the laser beam waist of ${\sim}30\;\mu$m. Solving the stability parameters using the higher-order approach, we can then solve the relation between the dc electric field gradient and the applied voltages using Eq.~\eqref{eq:gradE}. To explain the experimental results, a constant gradient must be included in the linear regression, yielding (in trap coordinates)
\begin{eqnarray} \label{eq:dEscan}
&&\begin{pmatrix} dE_x/dx \\ dE_y/dy \\ dE_z/dz \end{pmatrix} =
\left(\begin{array}{@{} S[table-format=-2.3] @{}}
-9.166 \\ -11.949 \\ 21.114 \end{array}\right) \frac{\mathrm{V}}{\mathrm{mm}^2} \nonumber \\  
&& + \left(\begin{array}{@{} S[table-format=-1.3]S[table-format=-1.3]S[table-format=-1.4]S[table-format=-1.4] @{}}
 -1.954 & -0.054 & -0.0415 & -0.0019 \\
 -1.961 & -0.049 & -0.0349 & -0.0114 \\
  3.915 &  0.104 &  0.0765 &  0.0133 \end{array}\right)
\frac{\mathbf{V}_\mathrm{bias}}{\mathrm{mm}^2},
\end{eqnarray}
where the bias voltage vector is $\mathbf{V}_\mathrm{bias} =  \left(V_{12,\mathrm{cm}}, V_{12,\mathrm{diff}}, V_{34,\mathrm{cm}}, V_{34,\mathrm{diff}}\right)\tran $.
The number of decimals reflects the fitting uncertainty.
The gradient created by the trap dc voltage $V_{12,\mathrm{cm}}$ gives the dc trap efficiency $\eta_U = 0.724$.
The measured gradients will be compared to simulations in Sec.~\ref{sec:FEM}.

\begin{figure}[tb]
\centering
\includegraphics[width=1\columnwidth]{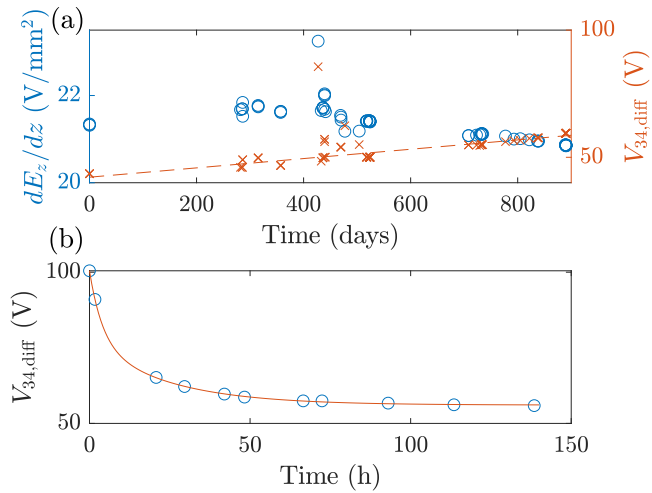}\\%
\caption{(a) Constant gradient (circles, left axis) and bias voltage $V_{34,\mathrm{diff}}$ (crosses, right axis) as functions of time. The dashed line is a fit to all voltage points within $\pm 3$\;V from the linear trend.
(b) Decay of $V_{34,\mathrm{diff}}$ (circles) with fitted double exponential decay curve (solid curve).
\label{fig:const-grad}}
\end{figure}

With the matrix in Eq.~\eqref{eq:dEscan} known, one can solve the constant gradient from all secular measurements separately. Figure~\ref{fig:const-grad}(a) shows the axial gradient as a function of time. Also shown is the bias voltage $V_{34,\mathrm{diff}}$ to demonstrate that abrupt changes in the gradient correlate with changes in the voltage needed to minimize the micromotion. Large changes typically occur when the imaging objective, which is very close to the vacuum chamber window, and the magnetic shields are removed and reinstalled and are likely due to charging of the window and/or objective lens. Figure~\ref{fig:const-grad}(b) shows how the voltage $V_{34,\mathrm{diff}}$ (that minimizes the micromotion) decays towards its prior value after such a charging has occurred. The decay is well described by a sum of two exponential decay functions with time constants of 4\;h and 24\;h. 

Figure~\ref{fig:const-grad}(a) also shows a linear increase in the baseline of $V_{34,\mathrm{diff}}$. This is believed to be due to irreversible positive charging of the antireflection coating of the fused-silica vacuum window caused by the 405-nm photo-ionization laser. \cite{Harlander2010a}
The voltage changes required for micromotion minimization following application of 405-nm light support this. The window on the opposite side is CaF$_2$ to allow thermal imaging of the trap for blackbody-radiation-shift evaluation, \cite{Dolezal2015a} which could explain the asymmetric effect. In contrast, using the oven is generally accompanied by a small change in the axial electric field, likely related to deposition of atoms and/or charges on the electrodes.

\subsection{Temperature dependence of the trap rf voltage \label{sec:T-dep}}

The measurement used to derive Eq.~\eqref{eq:dEscan} also yielded 52 $q_z$ values obtained over 12\;h. These are plotted in Fig.~\ref{fig:T-dep} together with the rectified monitor voltage $V_\mathrm{mon}$, see the \hyperref[sec:helres]{Appendix}, and the temperature $T_\mathrm{hel}$ of the helical resonator. There is a clear correlation between $q_z$ and $V_\mathrm{mon}$ and anti-correlation between $q_z$ and $T_\mathrm{hel}$. 

The resistivity temperature coefficient of copper is $4\times 10^{-3}\;\mathrm{K}^{-1}$. A simple electrical model of the helical resonator \cite{Siverns2012a} gives $q_z \propto V_0 \propto Q^{1/2} \propto R^{-1/2}$, where $Q$ and $R$ are the quality factor and resistance, respectively. This gives an expected temperature dependence of $-2\times 10^{-3}\;\mathrm{K}^{-1}$, whereas the linear fit of $q_z$ vs $T_\mathrm{hel}$ in Fig.~\ref{fig:T-dep}(c) yields $-2.9(4)\times 10^{-3}\;\mathrm{K}^{-1}$. The additional sensitivity could be due to temperature dependence of the impedance matching and resonance frequency of the helical resonator.

If one attempts to analyze this measurement using the lowest-order expression, Eq.~\eqref{eq:secular1}, a clear signature of failure is that the obtained $q_z$ values will depend linearly on $V_{12,\mathrm{cm}}$ to the degree that the correlation with $V_\mathrm{mon}$ and $T_\mathrm{hel}$ in Fig.~\ref{fig:T-dep} is lost.

\begin{figure}[tb]
\centering
\includegraphics[width=1\columnwidth]{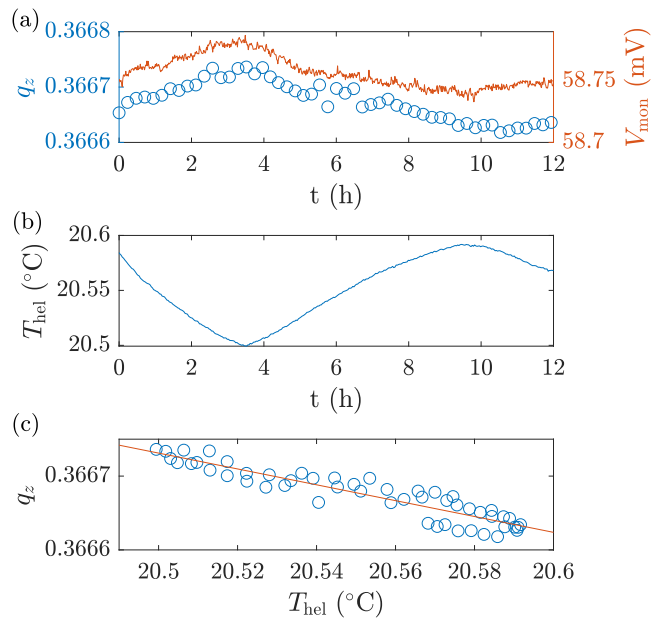}\\%
\caption{(a) Stability parameter $q_z$ (blue circles, left axis) and rectified monitor voltage $V_\mathrm{mon}$ (red line, right axis), and (b) helical resonator temperature $T_\mathrm{hel}$ as functions of time. (c) $q_z$ vs $T_\mathrm{hel}$ (blue circles) with linear fit (red line).
\label{fig:T-dep}}
\end{figure}

\section{Trap FEM simulation \label{sec:FEM}}

To assess the electric fields created by the applied trapping and bias voltages, an electrostatic finite-element-method (FEM) simulation of the trap was carried out. All the parts shown in Fig.~\ref{fig:trap} were included in the model. The main simplification was that the grounded vacuum chamber and magnetic shields outside the windows were replaced by a grounded boundary box approximating the inner dimensions of the former.

Starting with the rf field, the simulation shows that there is a slight `rf leakage' from the Cu trap body around the L-shaped, rf-grounded shield electrode supports, which displaces the rf minimum by $0.1\;\mu$m in the $+X$ direction. This point is taken as the origin of the chamber and trap coordinates.
The rf trap efficiency $\eta_V = 0.739$ is obtained from second-order fits to the simulated electric potential around the origin, separately along the $z$ axis and in the radial plane,  cf.\ Eq.~\eqref{eq:phi2}. The trap rf voltage has been estimated both using this trap efficiency and an electrical model of the helical resonator. \cite{Siverns2012a} The methods are in good agreement, indicating that the simulated $\eta_V$ is correct within ${\sim}0.03$, limited by the $Q$ factor uncertainty and experimental repeatability.

The dc fields are analyzed next.
Due to the L-shaped mounts of the shield electrodes, the quadrupole field created by these (with $V_3 = V_4 = 0$) is shifted by $8\;\mu$m in the $+X$ direction, which means that there is an electric field component $E_X>0$ at the origin. This field is cancelled when the micromotion is minimized, but, as will be seen, this results in a strong correlation between $V_{12,\mathrm{cm}}$ and $V_{34,\mathrm{cm}}$. The simulated trap efficiency of the shield electrodes is $\eta_U = 0.712$, slightly lower than that of the rf electrodes ($\eta_V = 0.739$). This is expected, as the shield electrodes are closer to the grounded oven assembly and vacuum chamber (grounded boundary box). The small discrepancy between the simulated and measured $\eta_U$ (0.712 vs 0.724) is likely due to the fabrication tolerances, as the trap efficiencies depend strongly on the exact electrode separations and the degree of roundover of the shield electrode apex.

Simulating the field created by all the dc electrodes, we can express the electric field at the origin (in chamber coordinates) as a function of the bias voltage vector as
\begin{equation} \label{eq:E0}
\begin{pmatrix} E_X \\ E_Y \\ E_Z \end{pmatrix} =
\left(\begin{array}{@{} S[table-format=-1.1]S[table-format=-3.1]S[table-format=-2.1]S[table-format=-1.1] @{}}
13.2 & 0.0 & -10.1 & 0.0 \\
-0.1 & 0.0 & 0.0 & -7.0 \\
0.0 & -167.6 & 0.0 & 0.0
\end{array}\right)
 \frac{\mathbf{V}_\mathrm{bias}}{\mathrm{m}}.
\end{equation}
As intended, the voltages $V_{34,\mathrm{cm}}$, $V_{34,\mathrm{diff}}$, and $V_{12,\mathrm{diff}}$ create compensation fields along the $X$, $Y$, and $Z$ axes, respectively. As mentioned above, due to the L-shaped shield electrode supports, $V_{12,\mathrm{cm}}$ creates a field along $x$ that needs to be cancelled using $V_{34,\mathrm{cm}}$. As the model is nominally mirror symmetric with respect to $Y$, the small $E_Y$ caused by $V_{12,\mathrm{cm}}$ is likely due to the numerical accuracy of the simulation.

From \eqref{eq:E0},  one can solve a voltage vector that does not change the field at the ion: $\Delta \mathbf{V}_\mathrm{bias} = \left(1, 0, 1.30, -0.03 \right)\tran $\;V. This can be used to change the axial-to-radial secular frequency ratio, governed by $V_{12,\mathrm{cm}}$, without increasing the EMM.
This vector has also been measured by varying $V_{12,\mathrm{cm}}$ and minimizing the micromotion at each setting. The two experimental results, $\Delta \mathbf{V}_\mathrm{bias} = \left(1, 0, 1.30, 0.02 \right)\tran$\;V and $\Delta \mathbf{V}_\mathrm{bias} = \left(1, 0, 1.30, -0.03 \right)\tran$\;V, are in excellent agreement with the simulated vector, indicating that the trim electrodes, whose tolerances are larger than those of the trap, are sufficiently well described by the FEM model. 

The simulated gradients at the origin are, in chamber coordinates,
\begin{equation} \label{eq:dE0sim}
\begin{pmatrix} dE_X/dX \\ dE_Y/dY \\ dE_Z/dZ \end{pmatrix} =
\left(\begin{array}{@{} S[table-format=-1.2] S[table-format=1] S[table-format=-1.3] S[table-format=1] @{}}
 -1.92 & 0 & -0.037 & 0 \\
 -1.92 & 0 & -0.030 & 0 \\
 3.85 & 0 & 0.067 & 0
\end{array}\right)
\frac{\mathbf{V}_\mathrm{bias}}{\mathrm{mm}^2}.
\end{equation}

For negligible radial rf asymmetry, the directions of the weaker ($x$) and stronger ($y$) radial trap axes can be found by finding the directions of maximum and minimum dc gradients in the radial plane, respectively. $V_{34,\mathrm{cm}}$ tends to align the trap axes with the chamber axes, while $V_{34,\mathrm{diff}}$ tends to align the trap axes at $45^\circ$ to these (this is why the gradient is zero in chamber coordinates). In this ideal model without patch potentials and external fields, the typical voltages $V_{34,\mathrm{cm}}=25$\;V and $V_{34,\mathrm{diff}}=55$\;V would align the weaker trap axis at $-51^\circ$ to the chamber $x$ axis.

In the experimentally determined trap coordinates, the gradients are
\begin{equation} \label{eq:dE0simtrap}
\begin{pmatrix} dE_x/dx \\ dE_y/dy \\ dE_z/dz \end{pmatrix} =
\left(\begin{array}{@{} S[table-format=-1.2] S[table-format=1] S[table-format=-1.3] S[table-format=-1.4] @{}}
 -1.92 & 0 & -0.037 & 0.003 \\
 -1.92 & 0 & -0.031 & -0.003 \\
 3.85 & 0 & 0.067 & 0.000
\end{array}\right)
\frac{\mathbf{V}_\mathrm{bias}}{\mathrm{mm}^2}.
\end{equation}
The only significant difference compared to chamber coordinates is that $V_{34,\mathrm{diff}}$ now creates a small gradient. Comparing Eq.~\eqref{eq:dE0simtrap} to the measured gradient, Eq.~\eqref{eq:dEscan}, the gradients created by $V_{12,\mathrm{cm}}$ and $V_{34,\mathrm{cm}}$ are in good agreement, with the measured values being slightly larger. The larger-than-simulated gradients created by the differentially applied voltages are attributed to manufacturing and assembly tolerances.

\section{Constant gradient \label{sec:gradEconst}}

The nearly constant gradient in Fig.~\ref{fig:const-grad}(a) and Eq.~\eqref{eq:dEscan} rotates the weak radial axis from the simulated $-51^\circ$ to the measured $-13^\circ$ angle. It is also believed to be largely responsible for the background field $(0.33, 0.40, 0.04)\;\mathrm{V/mm}$ (in chamber coordinates) that is compensated by the typically applied voltages, with the exception for the $Y$ component, which is (partly) caused by charging of the window as discussed in Sec.~\ref{sec:grad}.

Patch potentials can originate from contact potential (work function) variations on the electrode surfaces due to different crystal planes or adsorbed material.\cite{Wineland2002a} A contact potential caused by adsorption of atoms from the oven on the electrodes is a common concern. 

Patch potentials were simulated as floating $5\;\mu$m from the dc-grounded endcap electrode surfaces in order to electrically insulate them. It was found that slightly asymmetric patches covering ${\sim}80$\% of the endcap faces could cause a gradient of the correct shape. The required potential is, however, $-3.9$\;V, for which no physical explanation has been found. The work function of Sr is ${\sim}2.6$\;eV and that of Mo is $(4.36 \ldots 4.95)$\;eV depending on the crystal direction,\cite{Lide-CRC90} so Sr deposited on the electrodes should cause a positive potential of ${\sim} +2$\;V, similar to what has been observed for Mg deposited on Cu.\cite{Dholakia1993a}

Negative charging of insulating surfaces, including metals with a native oxide layer, when exposed to blue or ultra-violet light has been observed in many studies.\cite{Harlander2010a,Wang2011d,Allcock2012a,Harter2014a}  
However, these charges typically decay within minutes to days. 
We have not observed significant changes in the constant gradient or the micromotion compensation voltages even after a few months break in trapping ions or after having the same ion trapped for weeks without reloading.

An electric field gradient of similar strength has been observed near the loading region of a planar ion trap.\cite{Narayanan2011a} In this case, the gradient was accompanied by an increased anomalous heating rate,\cite{Daniilidis2011a} 
whereas our heating rate of $2.0(8)\;\mathrm{mK/s}$ is well in line with the values for other traps with polished Mo electrodes and photo-ionization loading when the electrode distance is accounted for.\cite{Dube2015a,Nisbet-Jones2016a} Thus the cause of the constant gradient remains unknown, but no negative effects related to it have been observed.


\section{Electric quadrupole shift prediction and cancellation \label{sec:EQS}}

The electric quadrupole shift (EQS) is caused by the interaction between the electric quadrupole moment of an atomic state, due to its charge distribution deviating from spherical symmetry, and an electric field gradient. For an ion with a $D_{5/2}$ clock state and no nuclear spin, it is usually written in the form \cite{Itano2000a,Shaniv2016a}
\begin{equation} \label{eq:EQStrad}
\Delta\nu_{\mathrm{EQS}} = -\frac{1}{4 h} \frac{d E_z}{d z} \Theta(D,\frac{5}{2}) \frac{12 m_{J'}^2 - 35}{40} f(\alpha,\beta,\epsilon_\mathrm{dc}),
\end{equation}
where $\Theta(D,\frac{5}{2}) = 2.973^{+0.026}_{-0.033} e a_0^2$ is the electric quadru\-pole moment, \cite{Shaniv2016a} $a_0$ is the Bohr radius, $m_{J'}$ is the $D_{5/2}$ magnetic quantum number, and the angular function is given by 
\begin{equation} \label{eq:EQSangle}
f(\alpha,\beta,\epsilon_\mathrm{dc}) = \left(3 \cos^2{\beta}-1\right) - \epsilon_\mathrm{dc} \sin^2{\beta} \cos{2\alpha}.
\end{equation}
Here $\epsilon_\mathrm{dc}$ is the radial asymmetry parameter of the dc quadrupole field and $\alpha$ and $\beta$ are Euler angles that take the frame defined by the principal axes $z$ of the quadrupole field to that defined by the quantization axis (magnetic field). \cite{Itano2000a} This form is convenient if the dc field is dominated by the applied trap dc potential, but not in the case of `arbitrary' $a_i$ parameters: $d E_z/dz$, $\epsilon_\mathrm{dc}$ and the rotation angles must be evaluated in the coordinates determined by the gradient principal axis, which can change as the trap and bias dc voltages are adjusted. In this case, if one defines the unit vector of the magnetic field in trap coordinates as $\mathbf{b}$, Eqs.~(\ref{eq:EQStrad}--\ref{eq:EQSangle}) can be rewritten in the following intuitive form,
\begin{equation} \label{eq:EQSsum}
\Delta\nu_{\mathrm{EQS}} = \frac{1}{4 h} \frac{m \Omega^2}{4 q} \Theta(D,\frac{5}{2}) \frac{12 m_{J'}^2 - 35}{40} 2 \sum_i a_i b_i^2,
\end{equation}
which is valid regardless of which $a_i$ component is the largest.

Zeeman averaging \cite{Dube2005a} is a powerful approach to cancel the EQS in, e.g., the $^{88}$Sr$^+$ and $^{40}$Ca$^+$ clocks. It utilizes the fact that $12 m_{J'}^2 - 35$ in Eqs.~\eqref{eq:EQStrad} and \eqref{eq:EQSsum} becomes zero when summed over $m_{J'} = 1/2, 3/2, 5/2$ or interpolated to $m_{J'}^2 = 35/12$. However, reducing the shift itself makes the evaluation of the cancellation level less stringent. \cite{Dube2013a} Minimizing the absolute shift can be even more important in clocks where the EQS is cancelled by averaging over three mutually orthogonal magnetic-field directions, \cite{Itano2000a} as the uncertainties of the field directions can be significant. This method is typically used in clocks utilizing a single $m_F = 0 \rightarrow m_{F'} = 0$ transition, e.g., $^{171}\mathrm{Yb}^+$ and $^{199}\mathrm{Hg}^+$.

From Eq.~\eqref{eq:EQSsum}, it is obvious that the EQS can be minimized in two ways: (i) by minimizing the $a_i$ parameters and (ii) by choosing the magnetic field direction appropriately. Minimizing the $a_i$ parameters first, decreases the sensitivity to drift in the background magnetic field. For example, by tuning $a_z$ to 0 using the trap dc voltage, we have $a_x = -a_y$ and the EQS can be nulled by choosing the magnetic field to have equal projections on the $x$ and $y$ axes. From Eq.~\eqref{eq:betaseries}, for a trap with radially symmetric rf field, $a_z = 0$ when the axial-to-radial secular frequency ratio is $\omega_z/\sqrt{(\omega_x^2 + \omega_y^2)/2} = 2 + (75/512) q_z^2 + \mathcal{O}(q_z^4)$.

\begin{figure}[tb]
\centering
\includegraphics[width=1\columnwidth]{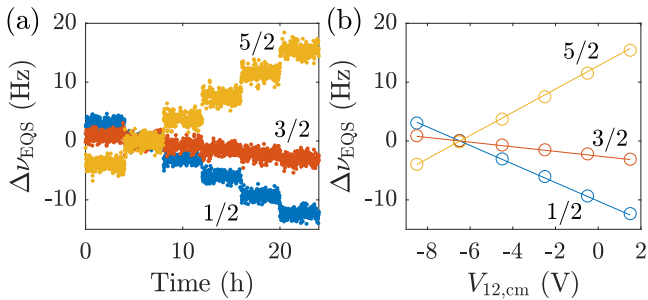}\\%
\caption{Measured EQS as a function of (a) time and (b) trap dc voltage together with theoretical predictions (solid lines). The labels refer to $|m_{J'}|$. The uncertainties of the experimental points in (b) are smaller than the symbols and error bars are omitted for clarity.
\label{fig:EQS}}
\end{figure}

Figure~\ref{fig:EQS} demonstrates the predictability of the EQS. 
Figure~\ref{fig:EQS}(a) plots the EQS of the different clock-state Zeeman sublevels from a measurement where $V_{12,\mathrm{cm}}$, and thus the EQS, were changed in discrete steps every 4\;h without increasing the EMM using the voltage vector $\Delta\mathbf{V}_\mathrm{bias}$.  Figure~\ref{fig:EQS}(b) shows the mean EQSs as a function of $V_{12,\mathrm{cm}}$ together with theoretical predictions calculated using Eqs.\ \eqref{eq:EQSsum}, \eqref{eq:gradE}, and \eqref{eq:dEscan}.  
The predictions were calculated using the constant gradient from a secular-frequency measurement 46 days earlier and the background magnetic field (${\sim} 1\;\mu$T) measured 4 months earlier. 
The direction of the total magnetic field (${\sim} 5\;\mu$T) was estimated using the measured relation between applied coil currents and magnetic field. 

For our experimental parameters, the ${\sim}10^{-5}$ accuracy of the $a_i$ parameters estimated in Sec.~\ref{sec:determine} corresponds to an EQS accuracy of ${\sim}10$\;mHz. Typical drifts in the background magnetic field affect the EQS at the same level. The difference between the measured and predicted shifts (${\lesssim} 0.3$\;Hz) in Fig.~\ref{fig:EQS}(b) is likely dominated by a change in the constant gradient, where a 1\% change corresponds to ${\sim}100$\;mHz in EQS.

The cancellation factor of the Zeeman-averaging EQS cancellation method 
has previously been estimated using clock servo simulations. \cite{Dube2013a} The measured EQS used as input in such a simulation contains quantum projection noise (QPN). Since we operate our clock with minimized EQS of the order of 10\;mHz, the QPN is two orders of magnitude larger than the EQS.

The predictability of the EQS and the possibility to switch voltages without increasing the EMM 
allowed a direct measurement of the EQS cancellation factor using two interleaved clock runs with different bias voltages $\mathbf{V}_\mathrm{bias}^\mathrm{low} = (-6.5, 0.2223, 23.8203, 58.1454)\tran$\;V and $\mathbf{V}_\mathrm{bias}^\mathrm{high} = (6.5, 0.2223, 40.7203, 58.1454)\tran$\;V, resulting in $\Delta\nu_{\mathrm{EQS}}(|m_{J'}| = 5/2) = -0.05(2)$\;Hz and $25.26(2)$\;Hz, respectively. To account for the time constant of the bias-voltage low-pass filters, a 100-ms delay was inserted after each switching of the voltages. The frequency difference between the interleaved low- and high-EQS clock runs averaged down as expected from QPN and after measuring for 17.7\;h, the difference was $0.0149(237)$\;Hz, consistent with zero within $1 \sigma$. A lower limit for the EQS cancellation factor can then be evaluated as the EQS difference between the two settings divided by the uncertainty of the difference measurement, which gives the factor 1070. As our clock typically is operated with $|\Delta\nu_{\mathrm{EQS}}| \lesssim 50$\;mHz, this is sufficient to reduce the EQS uncertainty to $\lesssim 1\times 10^{-19}$.


Using the vector $\Delta\mathbf{V}_\mathrm{bias}$, $V_{12,\mathrm{cm}}$ has been varied between $-7.5$\;V and $6.5$\;V without increasing the fractional scalar Stark and second-order Doppler shifts due to EMM to more than $\pm 5\times 10^{-19}$, respectively. At the magic frequency, this corresponds to a total EMM shift of $1.5\times 10^{-21}$. All other frequency shifts were common to the low and high EQS settings.

\section{Discussion and summary}

A method to accurately solve the Mathieu stability parameters $a_i$ and $q_i$ from measured secular frequencies was presented. It was then used to characterize an endcap trap. The gradients created by the applied voltages were shown to be in good agreement with a finite-element-method simulation. In addition, a constant gradient, which plays a significant role in defining the radial trap axes, was observed. For future trap design, we note that using two pairs of bias electrodes in the radial plane would allow adjusting the electric field and its gradient separately, offering control over the radial axes.

As an application for optical ion clocks, the electric quadrupole shift (EQS) was discussed. A simple relation between the EQS and the $a_i$ parameters was derived and it was shown that the measured relation between applied voltages and electric field gradients can be used to predict and minimize the EQS. Finally, a lower limit of 1070 for the cancellation factor of the Zeeman-averaging EQS cancellation method was obtained in an interleaved clock measurement.



%
%

%

\begin{acknowledgments}
T.L.\ thanks Pierre Dub\'e, Christian Tamm, and Tanja Mehlst\"aubler for helpful discussions. 
This work was supported by the projects 18SIB05 ROCIT and 20FUN01 TSCAC, which have
received funding from the EMPIR programme co-financed by the Participating States and from the European Union’s Horizon 2020 research and innovation programme.
The work was also supported by the Academy of Finland (REASON, decision 339821) and is part of the Academy of Finland Flagship Programme `Photonics Research and Innovation' (PREIN, decision 320168).
\end{acknowledgments}

\section*{Author Declarations}

\subsection*{Conflict of Interest}

The authors have no conflicts to disclose.

\section*{Data Availability}

The data that support the findings of this study are available from the corresponding author upon reasonable request.

\appendix

\section{Misaligned radial rf and dc axes \label{sec:diff-axes}}

If the radial axes $\tilde{x}, \tilde{y}$ of the rf potential are rotated around $z$ by the angle $\theta$ with respect to the dc axes $x,y$, the potential can be expressed as
\begin{equation} \label{eq:phiaqaxes}
\phi(\mathbf{x}) = \frac{m \Omega^2}{8 q} \sum_i a_i x_i^2 - 2q_i \cos{\Omega t} \tilde{x}_i^2,
\end{equation}
where $\tilde{x} = x \cos{\theta} + y \sin{\theta}$, $\tilde{y} = -x \sin{\theta} + y \cos{\theta}$, and $\tilde{z} = z$. Following \citet{Shaikh2012a}, but for a spherical Paul trap, the radial equations of motion in the dc coordinates become
\begin{subequations}
\begin{eqnarray} 
\frac{d^2 x}{d\tau^2}
+ a_x x - 2q_r\left[ \left( 1-\epsilon \cos{2\theta} \right) x 
+ \epsilon \sin{2\theta} y \right] \cos{2\tau} = 0, \nonumber \\
  \\
\frac{d^2 y}{d\tau^2} 
+ a_y y - 2q_r\left[ \epsilon \sin{2\theta} x 
+ \left( 1+\epsilon \cos{2\theta} \right) y \right] \cos{2\tau} = 0, \nonumber \\
  \label{eq:dummy}
\end{eqnarray}
\end{subequations}
where $q_r = -q_z/2$.
If $\epsilon=0$, $\theta=0$, or $a_x=a_y$, these decouple into Eq.~\eqref{eq:phiaq}; otherwise the equations have to be solved numerically.

Alternatively, one can use the pseudopotential approximation, where the pseudopotential energy of the ion is given by\cite{Dehmelt1967a,Wineland1983a}
\begin{equation} \label{eq:pp}
    \psi = \frac{q^2}{4 m \Omega^2} |\nabla \phi_\mathrm{rf}(t=0)|^2 + q \phi_\mathrm{dc},
\end{equation}
where $\phi_\mathrm{rf}$ and $\phi_\mathrm{dc}$ are the rf and dc parts of Eq.~\eqref{eq:phiaqaxes}, respectively. Taking the minimum and maximum of the radial pseudopotential as the effective radial trap axes $\hat{x}$ and $\hat{y}$, the angle $\varphi$ between these and the dc axes can be solved by substituting \eqref{eq:phiaqaxes} into \eqref{eq:pp}, transforming to cylindrical coordinates, and differentiating with respect to $\varphi$, which yields
\begin{equation} \label{eq:phi}
    \tan{2 \varphi} = \frac{(q_z^2/2) \epsilon \sin{2 \theta}}{\Delta a_{r} + (q_z^2/2) \epsilon \cos{2 \theta}}.
\end{equation}
Here $\Delta a_{r} = a_y-a_x$. As expected, $\Delta a_{r} = 0$ gives $\varphi = \theta$ while $\epsilon=0$ or $\theta=0$ gives $\varphi = 0$. If both the dc and rf asymmetry is significant, one can in principle determine the dc and rf axes by measuring the effective axes at different rf voltages ($q_z$) and extrapolating to zero and infinite $q_z$, respectively.

In the effective trap coordinates $\hat{x} \hat{y} \hat{z}$ defined by Eq.~\eqref{eq:phi}, the pseudopotential can be written as $\psi = \sum_i 
m \omega_{\hat{x}_i}^2 \hat{x}_i^2 /2$, where the radial secular frequencies are given by
\begin{eqnarray}
    \omega_{\hat{x},\hat{y}} &=& \frac{\Omega}{2} \bigg\{ \frac{a_x+a_y}{2} + \frac{q_z^2}{8}(1+\epsilon^2)  \nonumber \\ 
    && \mp \frac{1}{2} \left[\Delta a_r^2 + \Delta a_r q_z^2 \epsilon \cos{2\theta}  + \frac{q_z^4}{4} \epsilon^2 \right]^{1/2} \bigg\}^{1/2} \nonumber \\
    \label{eq:effsec}
\end{eqnarray}
and the axial frequency $\omega_{\hat{z}}$ by Eq.~\eqref{eq:secular1}. 
If $\epsilon=0$, $\theta=0$, or $\Delta a_r=0$, Eq.~\eqref{eq:effsec} reduces to Eq.~\eqref{eq:secular1} as expected. 
Since the pseudopotential approximation is of the same order as the lowest-order Mathieu solution, it has the same accuracy limitations. However, based on Eqs.~(\ref{eq:phi}--\ref{eq:effsec}), if $a_y-a_x \gg (q_z^2/2) \epsilon$, one can for practical purposes use the decoupled Mathieu equations and the higher-order method presented in Sec.~\ref{sec:determine}. This criterion is well fulfilled for the parameters measured in Sec.~\ref{sec:exp-param}.

\section{Tuneable helical resonator \label{sec:helres}}

\begin{figure}[tb]
\centering
\includegraphics[width=1\columnwidth]{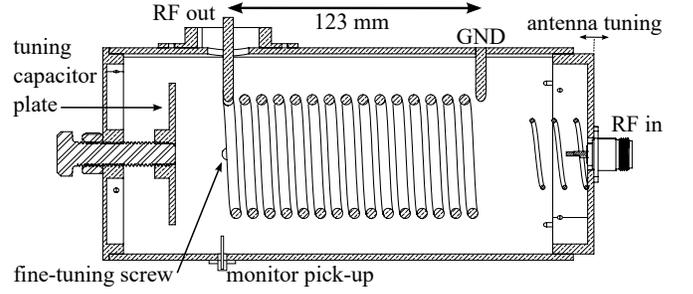}\\%
\caption{Helical resonator. The resonance frequency is tuned using the tuning capacitor (left) and fine-tuning screw. The resonator is impedance matched by tuning the distance between the antenna coil (right) and the main helical coil using the adjustable endcap. The capacitive pick-up antenna has a monitor ratio of $1:214$ when measured using a 10x oscilloscope probe. All parts are copper except for connectors and bolts/screws.
\label{fig:helres}}
\end{figure}

A drawing of the helical resonator is shown in Fig.~\ref{fig:helres}. In contrast to the common design, \cite{Macalpine1959a,Siverns2012a} the hot end of the helical rf coil exits the cylindrical shield through the side. This leaves the endcap free for a tuneable plate capacitor, which combined with a fine-tuning screw allows tuning the resonator to the magic frequency.

By measuring the resonance frequency $\omega_\mathrm{r}$ using multiple known capacitive loads and then the ion trap, the inductance $L$ and the capacitance $C_\mathrm{int}$ of the helical resonator secondary coil and the capacitance $C_\mathrm{trap}$ of the ion trap and feedthrough shield could be determined by fitting the function $\omega_\mathrm{r}^2 = [L(C_\mathrm{int}+C_\mathrm{ext})]^{-1}$, where $C_\mathrm{ext}$ is either the known capacitance or the trap capacitance. This gave $L=3.11\;\mu$H and $C_\mathrm{trap} = 29.2$\;pF, while the resonator capacitance $C_\mathrm{int}$ can be adjusted between 8.66\;pF and 10.87\;pF using the tuning capacitors, resulting in resonance frequencies of $14.26\ldots 14.67$\;MHz. The $Q$ value of the helical resonator connected to the trap is $830(30)$. A 5-V voltage amplitude from the signal generator into the impedance-matched helical resonator, corresponding to a power of 250\;mW, results in a trap voltage amplitude $V_0 = 350(20)$\;V.

Two capacitive pick-up antennas are used to monitor the rf voltage: one in the resonator shield, see Fig.~\ref{fig:helres}, and one in the grounded shield around the rf vacuum feedthrough. The one in the resonator shield is sensitive to small offsets in the position of the helical coil. Monitoring using an oscilloscope does not provide sufficient resolution to observe the fluctuations in the rf voltage caused by ambient temperature variations. The feedthrough monitor voltage is therefore rectified using a temperature-compensated diode rectifier similar to the one in \citet{Johnson2016a}, except that the input to the diode circuit is actively buffered with an operational amplifier. An example of this monitor signal is shown in Fig.~\ref{fig:T-dep}(a).


%

\end{document}